# Color-map recommendation for MR relaxometry maps

**Submitted to Magnetic Resonance in Medicine**

Miha Fuderer, Barbara Wichtmann, Fabio Crameri, Nandita M. deSouza, Bettina Baeßler, Vikas Gulani, Meiyun Wang, Dirk Poot, Ruud de Boer, Matt Cashmore, Wolter de Graaf, Kathryn E. Keenan, Dan Ma, Carolin Pirkl, Nico Sollmann, Sebastian Weingärtner, Stefano Mandija, Xavier Golay

Jointly, the authors form the Color Recommendation Committee (CRC)

## Abstract

Purpose: To harmonize the use of color for MR relaxometry maps and therefore recommend the use of specific color-maps for representing $T_1$ and $T_2$ maps.

Methods: Perceptually linearized color-maps were chosen to have similar color settings as those proposed by Griswold et al. in 2018. A Delphi process, polling the opinion of a panel of 81 experts, was used to generate consensus on the suitability of these maps.

Results: Consensus was reached on the suitability of the logarithm-processed Lipari color-map for $T_1$ and the logarithm-processed Navia color-map for $T_2$. There was consensus on color bars being mandatory and on the use of a specific value indicating "invalidity". There was no consensus on whether the ranges should be fixed per anatomy.

Conclusion: The authors recommend the use of the logarithm-processed Lipari color map for displaying quantitative $T_1$ maps and $R_1$ maps; likewise, the authors recommend the logarithm-processed Navia color-map for displaying $T_2$, $T_2^*$, $R_2$ and $R_2^*$ maps.

## Introduction

Magnetic resonance imaging (MRI) for clinical applications commonly uses qualitative $T_1$-weighted and $T_2$-weighted images, complemented by a plethora of other images based on different contrast mechanisms. Traditionally, these images are displayed in greyscale. Yet, MRI can also deliver quantifiable physical parameters including the relaxation time constants $T_1$ and $T_2$ and their inverses ($R_1$ and $R_2$), as well as related quantitifiable entities like $T_2^*$, $R_2^*$, $T_2'$ and $R_2'$. Mapping/imaging of these quantitative parameters dates back to at least the 1970s[1], but only gained traction in the last two decades, after acquisition techniques for quantification were optimized[2–5]. In the recent decade, this has included many types of transient-state sequences[6–10].

This paper focuses on quantitative relaxation maps. A *proton-density* map, which is often a by-product of generating relaxation maps, is not handled here, since there already is a *de facto* consensus to display proton density maps in greyscale. Opposed to this, there is no consensus yet on how to display relaxation maps.

Typically, quantitative maps (most notably, quantitative relaxation maps) are displayed in *color*. The use of color coding was first described by Pykett et al. for displaying the $T_1(x, y, z)$ distribution, which in turn provides a visual representation with greater contrast visibility than greyscale images[1,11].

In order to define in which way a scalar quantity is to be displayed in color, a *color-map* is needed. In this context, a *color-map* is a function that maps the scalar values of a range (e.g., from 0 to 100%, or from 0 to 255 for 8-bits storage, or from 20 ms to 50 ms) onto a path through a (3-dimensional) color space. With this definition, there is an infinite variety of possible color-maps.

In many clinical areas, quantitative relaxation maps have shown benefits over more conventionally used weighted images; an overview of clinical/translational applications is provided in a review by Tippareddy et al.[12]. There are also applications where a quantitative map is an intermediate product intended for synthesis (i.e. simulation) of "weighted" images[13]. In addition, qualitative images are very sensitive to imaging parameters leading to strong variations among systems and vendors; these variations often impede generalization of machine learning algorithms trained on one vendor's data to other imaging systems[14]. Quantitative maps, by their nature, could solve this issue, provided the measurements are sufficiently reproducible[15].

In quantitative imaging, there are several reasons for displaying images in color rather than in greyscale. First, a color is easier to remember, to compare, and to communicate than a greyscale level (Figure 1a1 and 1a2). Second, by systematically using one color-map for $T_1$ maps (Figure 1b1) and another, very different color-map for $T_2$ maps (Figure 1b2), the viewer can immediately recognize the image as a $T_1$ map, a $T_2$ map – or, alternatively, as a qualitative ("weighted") greyscale image. This characteristic is particularly relevant if one combines several types of images into one displayed image, as in Figure 1c: the color part displays a $T_2$-map of the cartilage, while the greyscale-displayed regions represent an image with mixed $T_1/T_2$-weighting, providing the anatomical context. The color immediately indicates the image type ($T_2$ map or anatomical context), even within one image. Furthermore, as observed by Pykett et al., for a given range, a color-map allows for more contrast visibility than a greyscale image.

Many of these benefits (e.g., comparing, communicating, or recognizing) can only be achieved if a standardized system of color-maps is used. Unfortunately, there exists a large variety of applied color-maps in recent literature on quantitative MR (Supplementary Material 1). In addition, many of the applied color-maps do not adhere to scientific standards, as postulated in the field of data visualization[16] as "perceptually uniform, perceptually ordered, color-vision deficiency and color-blind friendly, and readable in black and white prints".

To arrive at a community driven consensus, the Delphi method has previously been successful in a variety of fields. It has been initially applied for military purposes[17] and as a structured tool to forecast the (technological) future[18]. In the medical field, the Delphi method has been applied for a variety of purposes[19–24] and more specifically in imaging applications, like the recommendation of processes for renal MRI[25–27], MRI/CT/Ultrasound of small bowel and colon[28], prostate cancer MRI[29], and for lesion segmentation[30].

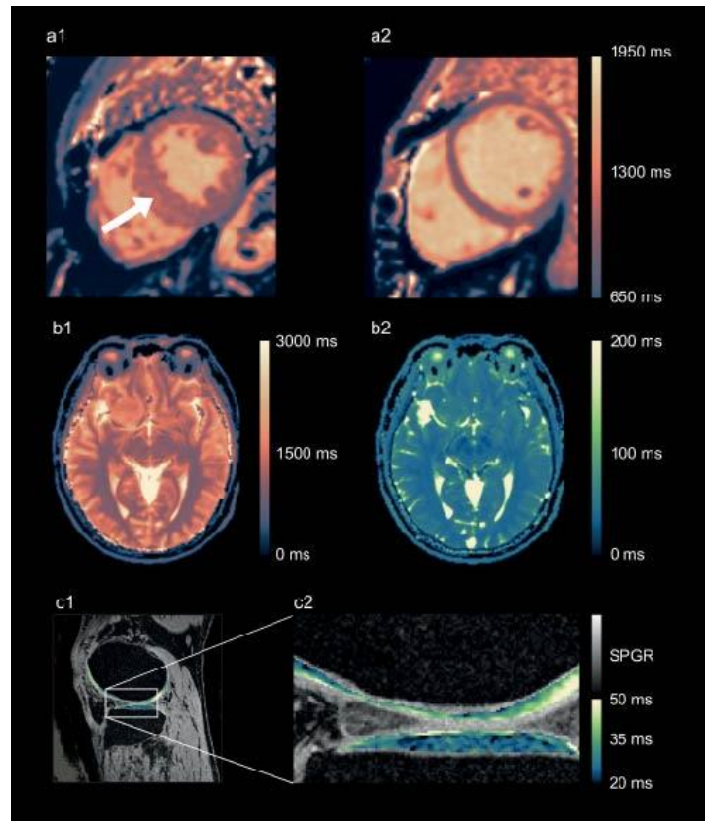

Figure 1: (a1 and a2) Maps of the myocardia of two different patients; healthy tissue (a2) may be depicted as purplish while an orange-like color points to pathology (a1, white arrow); (b1) example of a brain T1 map; (b2) example of a brain T2 map; (c1) an example of an image providing a T2 map of a knee (see c2 for detail): it contains a T2 map of cartilage (in color), as well as the context of "weighted" anatomy information (in grey).

To recommend and promote a well-motivated choice for the use of color-maps for the display of quantitative MR relaxation parameter maps, earlier work of Griswold et al.[31] was used as a starting point. A Delphi process was conducted to achieve a consensus recommendation for clinical adoption. The aim of the study was to provide distinct recommendations for color-maps designed for quantitative MRI relaxometry applications to be used in research as well as in the clinic.

# Methods

This effort originated with the quantitative MR Study Group (qMR-SG) of the International Society of Magnetic Resonance in Medicine (ISMRM), initiated by the first author (MF) and supported by members of the qMR-SG board (XG, BB, DM, MC). This group of five people recruited experts in the field of quantitative MR, using the following criteria: including both clinical and technical expertise, as well as experts on color-mapping, experts on the process, and industry representatives, while striving for a spread over geography, gender, and, particularly, balancing physicists and clinicians. This process led to a group of initially 16 people, the Color Recommendation Committee (CRC), which is composed of the authors of this publication, plus one representative of Canon and one representative of Philips, replacing Ruud de Boer after his retirement.

A Delphi process was used to develop a consensus recommendation for the display of color-maps. This process relied on a panel of experts who answer questionnaires in two or more rounds. After each round, a facilitator provided an anonymized summary of the experts' forecasts from the previous round as well as the reasons they provided for their judgments. Experts were encouraged to revise their earlier answers in light of the replies of other members of their panel.

## Defining the panel

A *panel* of 50-100 representative experts, which is central to the Delphi process, was selected as follows:

- Selecting 25 of the 100+ responders to the "original email", an informal email to the quantitative MR study group members in February 2022, polling the idea of a joint effort towards standardization of color-maps for relaxation. The selection was based on those responders who actively contributed to a discussion on the need to standardize color-maps (25 addressees).
- A random pick of 17 other respondents to the aforementioned email discussion.
- Key experts recommended by the CRC members (20 addressees).
- Adding one scientist on MRI known to be color-blind. (1 addressee).
- Requesting societies to suggest panel members, in order to obtain a better balance over anatomies (which contributed 8 addressees):
    - Quantitative Imaging Biomarkers Alliance
    - European Association of Cardiovascular Imaging
    - European Society of Radiology
    - Society of Abdominal Radiology
    - European Society of Gastrointestinal and Abdominal Radiology
    - Society for Advanced Body Imaging.

- Another round of suggestions by CRC members, specifically aimed at including more clinicians (9 addressees).
- All CRC members were part of the panel and answered the questionnaires. (16 addressees).

Of 96 addressees, 81 expressed willingness to cooperate (including the one person known to be color-blind); these 81 were defined as the panel.

Panel demographics were collected in Delphi round 2 (out of 4).

### Definition of questionnaires

A 9-point Likert scale was used in most questions, spanning the range from "fully disagree" to "fully agree". In general, the questionnaires went from generic to specific. Round 1 started with generic questions like establishing the number of required color-maps: should these be anatomy-specific? Should they differ between $T_1$ and $T_2$? (See Table 1 for the list of questions.) Furthermore, the relative importance of features like availability and perceptual linearity was probed.

One question specifically asked about the color-maps by Griswold et al.[31] which was motivated by the fact that Griswold's color-maps were designed with the same aim as outlined in the present recommendation (perceptually linear, with two distinct color-maps for $T_1$ and $T_2$). So, a shortcut was attempted by asking the opinion to the statement "The maps shown in Griswold 2018 should serve as a basis for the recommended color-maps". Should this have resulted in a clear agreement, then this would define the recommendation on color-maps.

In response to the comments of earlier rounds, some questions were added, re-formulated, grouped or split in subsequent rounds. For instance, in Round 2, the questions on the number of maps were reformulated. The question on the statement "The range (…) should be fixed" was split into "clinical" and "scientific" (see Table 2). Furthermore, the original three questions on perceptual linearity for different types of color-blindness were collapsed into one single question.

As an example of a comment-prompted question, Round 3 contained the question on the statement "Each quantitative relaxation image must be displayed in conjunction with a color-bar with adequately readable numbers."

As of Round 3, initial versions of the Lipari and Navia maps were presented, although the question was still open on whether $T_1$ and $T_2$ should get different maps.

In Round 4, the Lipari and Navia maps were further refined, based on the comments by the panel.

### Characteristics of the applied Delphi process

In most questions, the 9-points Likert scale was summarized to a 3-point Likert scale ("disagree" for scores 1, 2 and 3; "neutral" for scores 4, 5 and 6; "agree" for scores 7, 8 and 9), where, according to

Diamond et al.[32], a 75% consensus threshold was applied. This means that it was considered a consensus whenever the sum of scores 7, 8, 9 (or 1, 2, 3) reached 75% of all responses of that round. This *a-priori* set threshold of 75% agreement was used as it is a common choice in Delphi processes[32]. The questions on "importance" were exceptions, e.g., "The proposed color-maps should be as perceptually linear[16] as possible". For these questions, the 9-points score ranged from "Not important" to "Critical", and consensus was reached when the standard deviation of the scores was below 2.0, allowing, for example, a consensus on "moderately important".

The Delphi process in this study deviated from the conventional Delphi method, since the CRC occasionally modified questions from round to round. Regular CRC meetings took place during the process to discuss each round's outcomes and implement changes for subsequent rounds. These changes were motivated by several reasons. As an example, during the process, the comments of some panelists indicated that a question was worded ambiguously; alternatively, remarks pointed to omissions in the set of questions. In other cases, the set of possibilities was evolving from broad to narrow. Lastly, the suggested color-maps were adapted from round to round, based on the comments by the panel.

### Process behind the definition of the Lipari and Navia color-maps

To create the two scientific color-maps (Navia and Lipari; see ref. [33] for resources), five evenly-spread characteristic colors with variable lightness values were defined, to resemble the overall look of the two color-maps proposed in Griswold et al.[31]. To obtain a perceptually linear colormap based on the five anchor colors, the methodology of Crameri[34], as first outlined by Kovesi[35], was followed. These five colors were then complemented to a total of 256 individual colors following a smooth path within a perceptually uniform color space (here the L*a*b* color space according to the Commission Internationale de l' Eclairage[36], i.e. C.I.E.-L*a*b* space, or "CIELAB" in brief) passing these five chosen color values. In the following step, the perceptual differences between successive colors were calculated along the curve. From this, a cumulative sum of the perceptual differences along the color-map was formed, which was then divided into 256 equally spaced values, which were then mapped back onto the original color-space path via linear interpolation of the cumulative contrast curve. To form the final color-map, this procedure was done repeatedly until the variation of local contrast values along the curve came below a predetermined threshold. For practical purposes, the L*a*b* values are transformed to sRGB (standardized Red, Green, Blue)[37] values.

### Perceptual linearity – relative to the relaxation values or to the logarithm thereof?

The aforementioned process[34,35] delivers a color map that is perceptually linear. If applied to the relaxation maps themselves, then e.g. the difference between $T_2$=20ms and $T_2$=30ms becomes as conspicuous as the difference between, say 290ms and 300ms. The CRC questioned whether that is really the aim of the color map – or, whether according to the Weber-Fechner's law[38], the difference between 20ms and 30ms should be as conspicuous as the difference between 200ms and 300ms – i.e. whether it is more meaningful to make the color map perceptually linear with respect to the *logarithm* of the relaxation value. This question was presented to the panel. This is the rationale behind the "lin-log question" as entered in Delphi round 4. A possibility that was not considered at that time (see Discussion) was perceptual linearity with respect to $R_1$ or $R_2$.

Yet, "making the color map perceptually linear with respect to the logarithm of the value" does not involve taking the logarithm of the value maps. Rather, as outlined in Appendix 1, the color-map is stretched such that it becomes perceptually linear with respect to the logarithm. The appendix also explains that, in order to avoid log(0), it is not really a pure logarithm. In the sequel, the color-maps that are processed in this way are denoted as *logarithm-processed-Lipari* and *logarithm-processed-Navia*[33].

## 3 Results

### 3.1 Panel composition

The composition of the panel was assessed during Delphi round 2 and had the following characteristics:

- 22 medical professionals (45%) and 23 physicists (47%).
- 96% had a background in MRI, i.e. all but two – one of the being an expert on color visualization.
- Years of experience: ranges between 4 and 39, with an average of 19 years.
- Location (Figure 2): Europe (and particularly the Netherlands) was overrepresented (55%); China, India, and Japan were underrepresented (together, 6%), as well as Africa and Latin America (0%).
- One of the respondents (2%) had a type of color-blindness.
- Gender: 24% female (11), 76% male (37), 0% other.

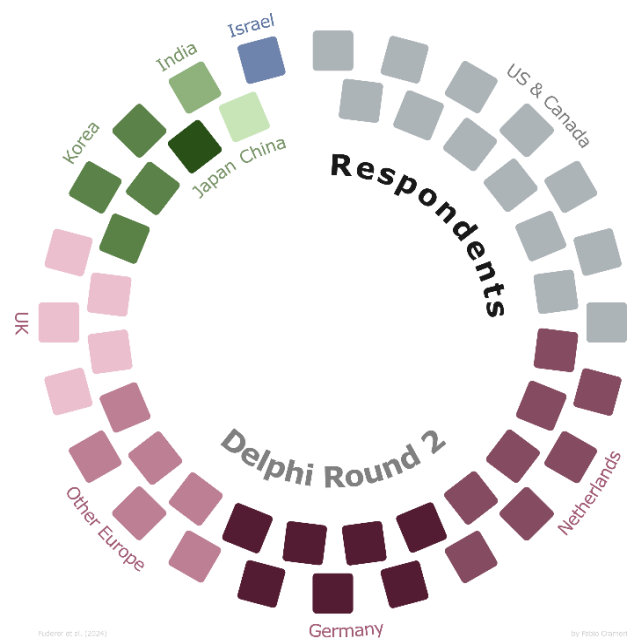

Figure 2: from Round 2, the number of respondents by country/region. Each block represents one person. Shades of purple reflect Europe, grey is North America, shades of green refer to Asia and blue refers to Middle East. One responder prefers not to state his region, so only 47 responders are shown here.

## 3.2 Delphi Round 1

The following items reached consensus during Round 1 (Table 1): the same color-map to be used for all anatomies, the need for a specific color to indicate invalidity, the choice of color scale and the software for the generation of color-maps should be available for free, and perceptual linearity is important – also considering people with the most common type of color-blindness (deuteranopia).

| Question (Number of respondents: 58 out of 81) | Score | Consensus? |
|---|---|---|
| One and the same color-map for (e.g.) T1 should be used for all anatomies | 78% agreed | yes |
| Within a given anatomy, it is a good idea to use a single color-map for all relaxation properties, including T1, T2, T2*, T2′, T1rho, R1, R2. So a T1 map may have a similar look to e.g. a T2 map. | 59% disagreed | no |
| Within a given anatomy, it is a good idea to define *exactly two* color-maps, i.e. one for all longitudinal-magnetization related relaxation (T1, R1) and one color-map for all of T1, T2, T2*, T2′, T1rho, R2 and R1rho. (If a relaxation rate map, e.g. R1, uses the reverse color-map of a relaxation time map, e.g. T1, this does not count as a different color-map) | 41% disagreed | no |
| Within a given anatomy, we need *more than two* different color-maps, e.g. one for each relaxation property, to represent all possible relaxation-related information. | 45% agreed | no |
| The range of an applied color-map should be in no way be fixed to certain relaxation values, but should always be freely adaptable. | 43% agreed | no |
| A color map should contain a specific color (e.g. black), clearly distinguishable from all the other colors of the color-map, to indicate invalidity (e.g. for regions of the image containing no nuclei and therefore having no, or unknown, relaxation properties). | 90% agreed | yes |
| Most (or all?) MR images are presented with a dark background. | 57% agreed | no |
| The maps shown in Griswold 2018 should serve as a basis for the recommended color-maps. | 38% agreed | no |
| The proposed color-maps should be as perceptually linear [Crameri 2020] as possible | 7.7±1.4 | yes, important |
| The proposed color-maps should be as perceptually linear as possible also when viewed by people with deuteranopia (most common red/green blindness; see [Crameri 2020]). | 6.9±1.8 | yes, fairly important |
| The proposed color-maps should be as perceptually linear as possible also when viewed by people with any kind of color blindness. | 6.2±2.1 | no (*) |
| The proposed color-maps should be as perceptually linear as possible also when converted to greyscale. | 6.7±2.3 | no (*) |
| The proposed color-maps should be available on common processing platforms like Pyplot and Matplotlib. | 7.5±2.3 | no (*) |
| The proposed color-maps should be available for free. | 8.1±1.0 | yes, very important |

Table 1: results from Round 1.
For the first eight items, agreement is reached if either agreement or disagreement exceeds 75%. Only the category coming closest to consensus is mentioned, which may be either the Likert categories 1, 2 and 3 or the Likert categories 7, 8 and 9. For the last six questions, asking on relative importance, the criterion is $\sigma < 2.0$. Consequently, the items marked (*) did not reach consensus because the standard deviation exceeded 2.0.

There was no consensus on the use of a single color scale for all relaxivity parameters, or whether two or more color scales would be needed for each relaxation property. Thus, the majority of items did not achieve consensus (Table 1), but the comments indicated that greater clarity was needed on range of colors and on perceptual linearity (Supplementary Material 2)

## 3.3 Delphi Round 2

Consensus was reached on the following points during Round 2: $R_1$ maps should have the same color scale (or inverse scale) as $T_1$ maps and color-maps should be as perceptually linear as possible – including *all* types of color-impairment. On the range to be applied for color-maps in the clinical applications, the responses came close to consensus; this did not apply to scientific work (Table 2). However, controversy remained around the display of $T_{1\rho}$ and whether $T_2$ and $T_2^*$ should be displayed similarly. There was also no consensus on whether the maps recommended by Griswold et al.[31] should serve as the basis of these recommendations.

| Question (Number of respondents: 48) | Score | Consensus? |
|---|---|---|
| T1 maps and T2 maps should get the same color-map | (a) | no |
| T2 maps and T2* maps should get the same color-map | 58% agreed | no |
| T2 maps and T1rho maps should get the same color-map | 44% agreed | no |
| Next to color-maps for T2 and T1rho map(s), there should be (an) additional color-map for T2-dispersion and/or T1rho-dispersion | (b) | no |
| R1 maps should get the same color-map as T1 maps, or the inverse thereof. | 79% agreed | yes |
| For clinical work, the range of an applied color-map should not be fixed to certain relaxation values, but should always be freely adaptable. | (a) | no |
| For clinical work, it is useful to have recommendations (per type of map and possibly per anatomy and field strength) on the range to be applied | 72% agreed | no |
| For scientific work, e.g. on the efficacy of obtaining quantitative maps, the range of an applied color-map should not be fixed to certain relaxation values, but should always be freely adaptable | 57% agreed | no |
| In current clinical practice, quantitative relaxation MR images are always read with dark background | 79% agreed | yes |
| The maps shown in Griswold2018 should serve as a basis for the recommended color-maps. | 47% agreed | no |
| The proposed color-maps should be as perceptually linear as possible, also when viewed by people with any kind of color blindness, i.e. perceptually linear when considering only the luminance component | 7.1±1.6 | yes, important |

Table 2: Results from Round 2
(a) On "T1 maps and T2 maps should get the same color-map", 54% disagreed, 38% agreed and there was almost no middle ground. Similarly for the question on freely adaptable ranges.
(b) On dispersion, 40% of panel-members scored a "5", interpreted as predominantly "no opinion"

## 3.4 Delphi Round 3

Initial versions of the Lipari and Navia color-maps were made available. This allowed presenting questions like "$T_1$ maps and $T_2$ maps should get the same color-map" in a different way, showing examples. The questions on $T_{1\rho}$ were omitted because too many panel members were indifferent, so there was no prospect of a consensus.

| Question (Number of respondents: 60) | Score | Consensus? |
|---|---|---|
| The [initial version of the] Lipari color-map is suitable for T1 maps. | 70% agreed | no |
| The [initial version of the] Navia color-map is suitable for T2 maps. | 53% agreed | no |
| Multiple choice between (a) Lipari for T1, Navia for T2; (b) Lipari for T1 and Lipari for T2; (c) grey for T1 and grey for T2. | (a) 72%<br>(b) 20%<br>(c) 8% | no |
| T2 maps and T2* maps should get the same color-map | 80% agreed | yes |
| Each quantitative relaxation image must be displayed in conjunction with a color-bar with adequately readable numbers. | 95% agreed | yes |
| For clinical work, it is useful to have recommendations (per type of map and possibly per anatomy and field strength) on the range to be applied | 73% agreed | no |

Table 3: Results from Round 3.

The question on adaptable vs. fixed minimum/maximum values for displaying the quantitative maps led to polarized opinions (see remarks in Supplementary Material 2) with no consensus in sight. Thus, this question was dropped and references to ranges were omitted from the recommendations.

The panel's comments pointed to an omission regarding the color bar, and as a result, a new question on the necessity of a color bar was added.

Results are shown in Table 3. An interesting result thereof is the multiple-choice question, where 72% chose the "Lipari for $T_1$, Navia for $T_2$" option, against 20% for having the same mapping for both. The 72% fell short of the preset threshold of 75% and a conventional Delphi process would necessitate repeating the question with the least popular choice (grey/grey) omitted. Yet, the CRC decided to treat the result as consensus, as the CRC deemed a future consensus very likely, and intended to avoid increased complexity in the 4[th] round.

### 3.5 Delphi Round 4

The responses from Round 3 led to further adaptations. The Lipari and Navia maps were improved according to the comments received. As explained in the Methods section, a question on logarithmic scale was entered here. Upon request, more examples were added, copies whereof are shown in Supplementary material 3; note that the colorbars in these examples were not yet in the recommended form, as explained in figure 3.

Two of the questions missed the threshold for consensus (Table 4): suitability of Navia met 72% agreement (36 out of 50) and the question about linear vs logarithmic scaling (Lin-Log) scored 70% (35 out of 50).

| Question<br>(50 respondents; now two of them expressed their color-blindness in a comment) | Score | Consensus? |
|---|---|---|
| The [improved version of the] Lipari color-map is suitable for T1 maps. | 75% agreed | yes |
| The [improved version of the] Navia color-map is suitable for T2 maps. | 72% agreed | no |
| [New] The logarithmic maps are at least as suitable as linear maps. | 70% agreed | no |

Table 4: Results from Round 4

When considering the Lin-Log question, some of the respondents were confused by the wording of the question. Others were in favor of the "Log" images and of the principle behind it, but objected to the presentation of the color bar. (In the presented examples, the color bar in "Log" looked the same as in "Lin", but the legend numbers were logarithmically spaced, Figure 3a.) As a result, the presentation of the color bar and its annotation was revised (Figure 3b) and the question was reformulated. Then, the revision was not sent to the full panel, but only to those 8 respondents who were not unequivocally in favor of the logarithmic scaling and who identified themselves in the questionnaire. This resulted in 3 respondents in favor of a logarithmic scale and 5 in favor of a linear one. These 3 were added to the 35 respondents who approved the logarithmic scale in round 4, bringing the total to 38, i.e. 76%.

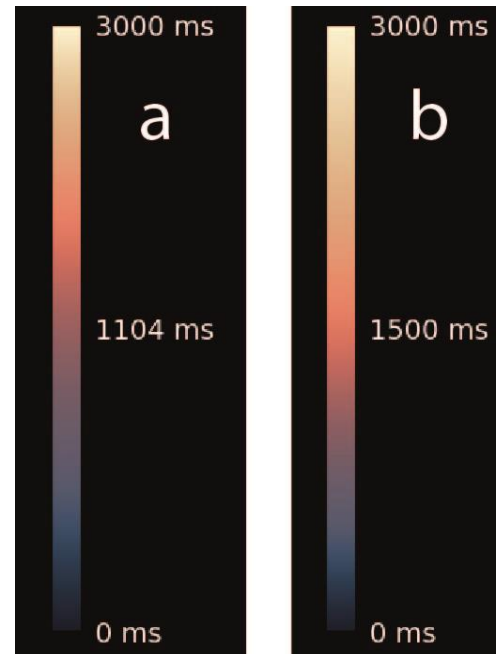

Figure 3: The colorbar and labeling as used (a) in the questionnaire of Round 4 and (b) as used in the last, limited, questionnaire. The images were identical in both cases, e.g., 1500 ms being displayed as orange in both cases. (b) is the recommended choice.

### 3.5 Summary of Delphi outcomes

Together, the following outcomes reached consensus:

- The color-map should be independent of anatomy.
- A color-map should contain a specific color (e.g., black), clearly distinguishable from all the other colors of the color-map, to indicate invalidity (e.g., for regions of the image containing no tissue and therefore having no, or unknown, relaxation properties).
- $T_2$ maps and $T_2^*$ maps should get the same color-map.
- $R_1$ maps should get the same color-map as $T_1$ maps, or the inverse thereof (an ambiguous wording, leaving two options open).
- Each quantitative relaxation image must be displayed in conjunction with a color bar with adequately readable numbers.
- The designed color-maps, Lipari and Navia, have to be available for free and have to be as perceptually linear as possible, also when viewed by people with any kind of color-blindness (i.e., perceptually linear when considering only the luminance component).
- Lipari is suitable for $T_1$, Navia is suitable (*) for $T_2$.
- The color-maps should be perceptually linear relative to the *logarithm* of the relaxation value.

(*) Navia for $T_2$ was just below consensus threshold, but the CRC expected consensus in the next round.

# Discussion

## Scope of this paper
While this paper focuses on relaxation times, MRI can be used to generate many other quantitative imaging biomarkers. This Delphi-study on relaxation-related color-maps can serve as an example that can be applied to other quantitative maps in future work.

## The impact of color on contrast-visibility
A benefit of using color is that it allows for more (color-)contrast visibility than a greyscale image. This has been quantitatively analyzed, using the $\Delta E_{CIEDE}$ metric[16], which expresses the perceptual contrast between neighboring values. This metric showed that the Lipari and Navia were superior to a greyscale-map. Lipari and Navia scored an average $\Delta E_{CIEDE}$ of around 0.44 against 0.29 for a greyscale-map. Although a color-map like Jet may have a still higher average score, this comes at the cost of locally over-enhancing or suppressing contrasts, and at the cost of being non-monotonic in luminance (the brightest color, yellow, being halfway the range), therefore not easily interpretable by color-impaired viewers.

## Displaying the units
Some details of the recommendation were retrospectively added by the CRC. In round 3, the following statement reached clear consensus: "Each quantitative relaxation image must be displayed in conjunction with a color-bar with adequately readable numbers". In retrospect, the CRC realized that the intended statement should have been "Each quantitative relaxation image must be displayed in conjunction with a color-bar with adequately readable numbers *and units*". Although the addition *"and units"* is strictly speaking not a Delphi outcome, the CRC has sufficient confidence that such addition would not have altered the consensus. This addition is reflected in the summary and the conclusions.

## Shortcuts used after Round 4
After Round 4, a small-scale questionnaire on the Lin-Log question was submitted to 8 participants because the declining number of responders with each round indicated respondent-fatigue and did not justify a subsequent "full" repeat round. Therefore, the extra round was limited to all non-anonymous participants who had not agreed to logarithmic maps in Round 4 and who had indicated they were unclear about the question being asked. In this extra round on clarifying the question, 3 of the 8 responded positively to the logarithmic maps, taking the total score of those preferring logarithmic maps to 76%.

For the same reason of responder-fatigue, no further iterations on the Navia color-map were done after Round 4. In Round 4, agreement to Navia scored 36/50 (or 72%); only two panel members disagreed (the remaining 12 were indifferent). One of these two was a respondent who systematically indicated – against all the others –a lack of interest in display of a quantitative map but only an interest in *interpretation* of the quantitative map (with color-labeled "normal" or "abnormal"). The case for Navia was further strengthened by two color-blind panel members in round 4, who were in favor of the Navia color-map. Given this situation, Navia was accepted as a consensus even though it formally achieved a 72% agreement.

Finally, the last question ("useful to have recommendations (…) on the range") was also very close to consensus (73%), so it would very likely result in full consensus in the next round. However, during the discussion, it was realized that recommendations about ranges, to be useful, needed to be set on an organ by organ basis, with a very clear steer from specialists in every single application of quantitative relaxometry. As the CRC was composed of a sub-group of such specialists only, it was decided not to pursue this recommendation for the time being. While range recommendations are relevant, it remains future work.

## Community-submitted discussion points

The discussion points below were entered by attendants to the qMR study group meeting at the ISMRM 2023 or these were submitted as part of the endorsement process.

- One panel member systematically rejected the proposed color-maps; in this person's view, color should be used to indicate normal vs. abnormal tissue. This idea is interpreted as a suggestion on a color-map for an interpretation or segmentation of the tissue (which may be based on quantitative maps), but not as a color-map for the quantitative ($T_1$ or $T_2$) maps themselves.
- $R_1$ (or $R_2$) relate to physical quantities like contrast uptake, so it would be most logical to make a color-map perceptually linear with respect to $1/T_1$, rather than $\log(T_1)$. While this is a valid argument, it was only recognized after closure of the Delphi process. Furthermore, the difference between the two curves would be slight, since both for $1/T_1$ and for $\log(T_1)$, a deviation from the theoretical curve is required for small $T_1$ values (see Appendix 1). Further modifications to the current Delphi-outcome are considered as future work.
- Perceptual linearity is, in principle, guaranteed in the sRGB color space. I.e., the color-maps are perceptually linear for a default display device as defined by the CIE[37] – i.e., for the "average" display system. In practice, no two display devices are truly identical, particularly considering (mis-)adjustments, and (for liquid-crystal displays) viewing-angle dependencies.
- The "invalid" color has been chosen as black. This has been hinted at in the Delphi process statement "A color-map should contain a specific color (e.g. black), clearly distinguishable from all the other colors of the color-map, to indicate invalidity", but the "e.g." still leaves some room for choices. Yet, any monochromatic choice, other than black or white, could be seen by color-impaired people as one of the "valid" colors. A pattern (e.g. a checkerboard) would indicate invalidity more clearly, but it would complicate the viewing pipeline – and it would not conform with the statement presented in the Delphi process. Between black and white, the CRC chooses for black, since it causes less glare. Care has been taken that the first valid color is perceptually at least 10% distant from black.

### Future work

While consensus was achieved on many points, the divergence of opinions amongst experts leaves a number of issues unresolved. In particular, the question on "The range of an applied color-map should [not] be fixed" (in jargon: allow window/level settings) explicitly did not reach consensus, with clear argumentation by proponents and by opponents on each side. Its alternative, "it is useful to have recommendations (…) on the range" came close to consensus; yet, as discussed, it was decided not to pursue this recommendation for the time being. While range recommendations are relevant, it remains future work.

Another question, which reached consensus but remained ambiguous was worded as "$R_1$ maps should get the same color-map as $T_1$ maps, or the inverse thereof". The remaining ambiguity was never resolved. As a consequence, it is recommended to use Lipari for $R_1$ and Navia for $R_2$, but it is explicitly left open on whether the maps should be reversed when applying these to $R_1$ or $R_2$. Both choices have proponents: on one hand, it is intuitive to systematically associate high values to bright image regions; on the other, a reversed map would mean that one gets almost the same image when applying a logarithmic display of a $T_1$ map compared to the inverted logarithmic display of the associated $R_1$ map. The latter is shown in Figure 4.

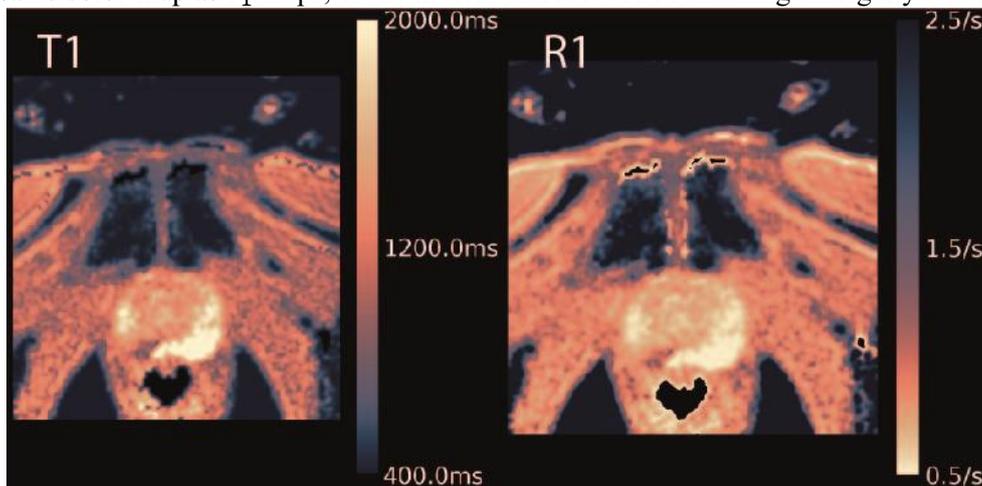

Figure 4: Example of a prostate T1 map (left) and the corresponding R1 map (right) displayed using the inverted Lipari color-map. Given the log-like function, with the inverted map, the R1 looks almost identical to the T1 map. Note that the logarithm is not applied to the maps themselves, but by processing the color-map. This is apparent by the difference between the colorbars: the color corresponding to the $T_1$ value that is linearly halfway 400ms and 2000ms is much brighter than the color corresponding to the $R_1$ value that is linearly halfway 0.5/s and 2.5/s.

The polling on $T_{1\rho}$ was dropped from the Delphi questionnaire after the 2nd round, as the responses indicated indifference of the panel members. This indifference may be partly related to the fact that the term $T_{1\rho}$ does not describe a scalar value but that it actually represents a (lock-field-dependent) continuum of values for any single tissue[39]. Thus, to reach a consensus on $T_{1\rho}$, a detailed discussion, providing the context of use and a rigorous physical definition, may be needed; this is beyond the scope of the present work.

> **Resulting recommendations**
>
> **The logarithm-processed Lipari color-map should be used for $T_1$ maps and the logarithm-processed Navia color-map should be used for $T_2$ and $T_2^*$. The color-maps and the logarithm-processing are jointly available in <https://doi.org/10.5281/zenodo.11082098>. The same color-maps are to be applied on $R_1$ and $R_2$ and $R_2^*$ respectively. This recommendation holds for all anatomies. The value of 0 is to be used to indicate that the calculated relaxation value is unknown (or "invalid") at that specific pixel, and it always should map to black.**
>
> **In addition, each quantitative relaxation image should be displayed in conjunction with a color bar with adequately readable numbers and units.**
>
> **These recommendations, which apply both to commercial display systems as well as to scientific publications, only achieve the aimed benefits with wide adoption. From this follows a plea on industry to adopt the recommendations, on the scientific community to use these recommendations when internally communicating MR relaxation maps – and *a fortiori* when publishing these. Similarly, colleagues are encouraged to promote the use of these recommendations, e.g., when peer-reviewing.**

# Conclusion

As outlined in the resulting recommendations, the CRC and the endorsers are confident to recommend the logarithm-processed Lipari color-map for $T_1$ maps, and the logarithm-processed Navia color-map for $T_2$-like maps. Combined with the recommendation that each quantitative relaxation image has to be displayed in conjunction with a color bar, this will lead to more uniformity, more comparability across results and easier recognizability of the map type.

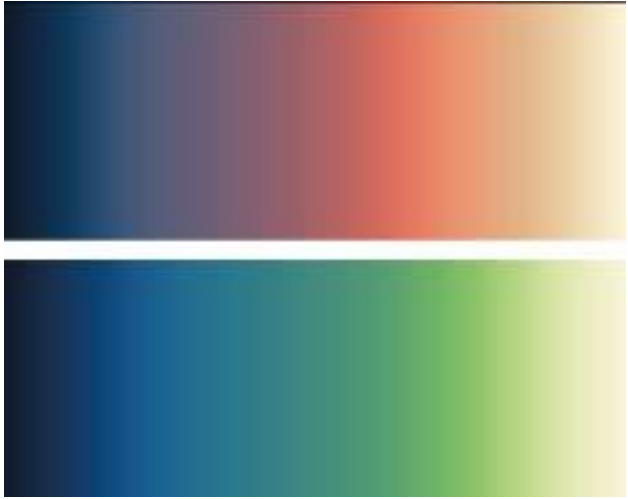

Figure 5: The recommended (unprocessed) color maps: Lipari (top) and Navia (bottom)

See figure 5 for the unprocessed maps.

# Acknowledgements

The authors would like to thank all panel members for their contribution. In particular, the authors would also like to thank the contributions by Sam Adams, Esha Baidya, Matthew Birkbeck, Lucie Chalet, Kay Chioma Igwe, Patricia Clement, Gastao Cruz, Yiyun Dong, Luke Edwards, Ruud B. van Heeswijk, Juan A. Hernandez-Tamames, Haotian Hong, H. Harry Hu, Dimitrios Karampinos, Liam Lawrence, Namgyun Lee, Maarten Naeyaert, Martijn Nagtegaal, Sophie Schauman, Marion Smits, Dennis C. Thomas, Ceren Tozlu and Marta Zerunian for suggesting improvements to the manuscript.

Likewise, the authors would like to thank Thomas A. Treibel for the cardiac image examples and Edwin Oei for the knee images.

The authors are also grateful to all endorsers of this paper.

The contribution of the first author has been financed by NWO grant number 17986

# Appendix 1: Details of the "logarithm" lookup

The recommendation is to map these color ranges to the *logarithm* of $T_1$ or $T_2$ values, rather than to the $T_1$ or $T_2$ values themselves.

This, however, does not mean that an explicitly logarithm has to be applied; rather, the proposal is to process the Lipari or Navia maps in such a way that their processed versions become perceptually linear to the logarithm of the input rather than being perceptually linear to the input (see figure 3b). As there are no recommendations on the ranges of $T_1$ or $T_2$, the lower level of the scale may be zero (which makes the logarithm ill-defined) or even negative. A negative lower-level makes no sense if the display device is well-calibrated and the color-map is perceptually linear; but if this ideal situation is not met, some users may be tempted to set the lower-level to a negative value – and most user-interfaces do allow it.

For that purpose, a parametrized function is defined having a linear segment and a logarithmic segment. In the following, it is assumed that:

- A pixel with a value of zero represents "invalid/unknown", i.e. the quantitative value is unknown at that pixel.
- The values of $L$ and $U$ represent the lower and upper end of the range of values to be displayed. For $U$ it is assumed that it is strictly positive; for $L$, explicitly no such assumption holds.
- The value of $\epsilon$ represents the smallest value that represents a meaningful relaxation value. The relaxation value that this may refer to, may depend on the rescale slope (in DICOM terms); e.g. the stored pixel values may range from 0 to 255, but actually represent $T_1$ values ranging from 0 to 2000 ms. In that example, the rescale slope is 7.843 and $\epsilon$ (which corresponds to a stored value of 1) equals 7.843.
- The output $f(x)$ ranges from 0 to 1, which can be scaled arbitrarily to the length of the color-map (e.g., by multiplying by 511.99 for a 512-entry table). Here, $f(x) = 0$ defines the "invalid" color (i.e., black). Here, $\epsilon_f$ refers to the value of $f$ that maps to the smallest nonzero color entry (in the example above, to 1/511).
- The auxiliary variable $a$, which separates the linear part from a logarithmic part, is declared as $a = U \cdot exp\,(-1)$
- and the auxiliary variable $b$, which equals $f(x)$ for $x = \max(a, L)$,
  is calculated as $b = \begin{cases} \frac{a-L}{2a-L} + \epsilon_f & a \geq L \\ \epsilon_f & a \leq L \end{cases}$

Herewith,

$$f(x) = \begin{cases} 1 & U \leq x \\ \dfrac{ln\left(\dfrac{x}{max(a,L)}\right)}{ln\left(\dfrac{U}{max(a,L)}\right)}(1-b) + b & max(a,L) \leq x \leq U \\ \dfrac{x-L}{(a-L)}(b - \epsilon_f) + \epsilon_f & max(0,L) < x \leq a \quad if \quad L < a \\ \epsilon_f & 0 < x \leq L \quad if \quad 0 < L \\ 0 & x = 0 \end{cases}$$

The Lipari and Navia maps themselves are publicly available[40]. The tooling to apply the logarithm-processing, and to apply the results to images, is available in Matlab or Julia[33].

# Figure captions

- Figure 1: (a1 and a2) Maps of the myocardia of two different patients; healthy tissue (a2) may be depicted as purplish while an orange-like color points to pathology (a1, white arrow); (b1) example of a brain T1 map; (b2) example of a brain T2 map; (c1) an example of an image providing a T2 map of a knee (see c2 for detail): it contains a T2 map of cartilage (in color), as well as the context of "weighted" anatomy information (in grey).
- Figure 2: from Round 2, the number of respondents by country/region. Each block represents one person. Shades of purple reflect Europe, grey is North America, shades of green refer to Asia and blue refers to Middle East. One responder prefer not to state his region, so only 47 responders are shown here.
- Figure 3: The colorbar and labeling as used (a) in the questionnaire of Round 4 and (b) as used in the last, limited, questionnaire. The images were identical in both cases, e.g., 1500 ms being displayed as orange in both cases. (b) is the recommended choice.
- Figure 4: Example of a prostate T1 map (left) and the corresponding R1 map (right) displayed using the logarithm-processed inverted Lipari color-map. Given the logarithm-processing, with the inverted map, the R1 looks almost identical to the T1 map. Note that the logarithm is not applied to the maps themselves, but by processing the color-map. This is apparent by the difference between the colorbars: the color corresponding to the $T_1$ value that is linearly halfway 400ms and 2000ms is much brighter than the color corresponding to the $R_1$ value that is linearly halfway 0.5/s and 2.5/s.
- Figure 5: The recommended (unprocessed) color maps: Lipari (top) and Navia (bottom).

# Table captions

- Table 1: results from Round 1.
  For the first eight items, agreement is reached if either agreement or disagreement exceeds 75%. Only the category coming closest to consensus is mentioned, which may be either the Likert categories 1, 2 and 3 or the Likert categories 7, 8 and 9. For the last six questions, asking on relative importance, the criterion is $\sigma < 2.0$. Consequently, the items marked (*) did not reach consensus because the standard deviation exceeded 2.0.
- Table 2: Results from Round 2
  (a) On "T1 maps and T2 maps should get the same color-map", 54% disagreed, 38% agreed and there was almost no middle ground. Similarly for the question on freely adaptable ranges.
  (b) On dispersion, 40% of panel-members scored a "5", interpreted by us as predominantly "no opinion"

- Table 3: Results from round 3
- Table 4: Results from round 4
- Table 5: Resulting recommendations

# Supplementary material

- Supplementary 1: The variety of color-maps used in current literature on relaxation.
- Supplementary 2: Collection of received comments.
- Supplementary 3: The images as presented in the questionnaire of round 4 of the Delphi process.
- Supplementary 4: [To be provided after the endorsement process] The list of endorsers.

Note that supplementary material is not provided on arXiv. Contact first author for information

# Supplementary material, not for publication

[Pending] Approval by the publication committee of the ISMRM